\documentclass[sigconf]{acmart}
\usepackage{enumitem}
\usepackage{tightenum}
\usepackage{listings}
\usepackage{tabularx}
\usepackage{diagbox}
\usepackage{setspace}
\usepackage{epsfig}

\usepackage{url}
\usepackage{subcaption}
\usepackage[ruled]{algorithm2e}

\usepackage{booktabs}
\usepackage{multirow}
\usepackage{makecell}
\usepackage{tikz}
\usepackage{wrapfig}
\usepackage{float}

\newcommand{\ignore}[1]{}

\newcommand{\beforecaption}{\vspace{-.15cm}\begin{spacing}{0.85}}
\newcommand{\aftercaption}{\vspace{-.15cm}\end{spacing}}

\usepackage{algorithmic}
\usepackage{graphicx}
\usepackage{textcomp}
\usepackage{xcolor}
\def\BibTeX{{\rm B\kern-.05em{\sc i\kern-.025em b}\kern-.08em
    T\kern-.1667em\lower.7ex\hbox{E}\kern-.125emX}}

\usepackage{multirow}
\usepackage{csquotes}
\newcommand{\Name}{LLmiRQ}

\usepackage{mdwlist}

 \usepackage{balance}

\usepackage[framemethod=TikZ]{mdframed}

\definecolor{Light}{gray}{.85}
\newcounter{findingsCounter}
\usepackage{tcolorbox}
\begin{document}

\title{Enhancing IR-based Fault Localization using Large Language Models}

\author{Shuai Shao}
\affiliation{\institution{University of Connecticut}
\city{Storrs, CT}
\country{USA}}

\author{Tingting Yu}
\affiliation{\institution{University of Connecticut}
\city{Storrs, CT}
\country{USA}}

\begin{abstract}
Information Retrieval-based Fault Localization (IRFL) techniques aim to identify source files containing the root causes of reported failures. While existing techniques excel in ranking source files, challenges persist in bug report analysis and query construction, leading to potential information loss. Leveraging large language models like GPT-4, this paper enhances IRFL by categorizing bug reports based on programming entities, stack traces, and natural language text. Tailored query strategies, the initial step in our approach (LLmiRQ), are applied to each category. To address inaccuracies in queries, we introduce a user and conversational-based query reformulation approach, termed LLmiRQ+. 
Additionally, to further enhance query utilization, we implement a learning-to-rank model that leverages key features such as class name match score and call graph score. This approach significantly improves the relevance and accuracy of queries.
Evaluation on 46 projects with 6,340 bug reports yields an MRR of 0.6770 and MAP of 0.5118, surpassing seven state-of-the-art IRFL techniques, showcasing superior performance.
\end{abstract}

\maketitle




\section{Introduction}
Fault localization plays a crucial role in software maintenance and development processes. When a bug is discovered by a user or a software developer, it is typically reported in a bug tracking system, such as Bugzilla~\cite{bugzilla}, Google Code Issue Tracker~\cite{gooledoclink}, or Github Issue Tracker~\cite{githublink}, through a bug report or an issue report. These reports offer valuable information that assists developers in efficiently identifying and resolving bugs.

Upon receiving a bug report, a developer's primary tasks include reproducing and localizing the bug within the source code. If we consider the bug report as a query and the source code files in the software repository as a collection of documents, the challenge of identifying relevant source files for a given bug report aligns with a standard task in information retrieval (IR). Several approaches have been proposed in IR-based fault localization~\cite{buglocator, brtracer, amalgam, bluir}, where similarity scores between bug reports and program entities (e.g., source files) are computed. Subsequently, rankings derived from these similarity scores reveal the relevance of software entities to the specific bug report.

While IR-based fault localization (IRFL) has undergone extensive research and development, it encounters significant challenges. A recent study by Lee et al.~\cite{bench4bl} assessed the reproducibility of IRFL performance. Despite employing various retrieval methods, resources, and query formulation strategies in six investigated techniques—BugLocator, BLUiR, BRTracer, AmaLgam, BLIA, and Locus—no significant performance differences were observed.

A key challenge in IRFL lies in constructing effective queries for retrieval. Common approaches view bug reports as queries, often involving tokenization and stop word removal~\cite{blizzard, brtracer, blia, bluir}. However, these methods frequently retain noise within the query. Additionally, existing query formulation strategies assign different weights to bug report components, such as the title, description, and steps to reproduce, for query construction~\cite{chaparro2019using, chaparro2017using}. Some strategies leverage graph creation and term weighting to identify essential tokens for query construction~\cite{blizzard, rahman2018effective}. Despite these efforts, these methods may lack a holistic understanding of the semantic content in bug reports, leading to noise introduction or critical information oversight.

The emergence of Large Language Models (LLMs)~\cite{vaswani2017attention, brown2020language} like ChatGPT has revolutionized natural language understanding. LLMs excel in interpreting natural language nuances and contexts, offering a promising avenue for deeper bug report analysis and construction of effective queries to enhance IRFL performance.

This paper introduces \Name{}, an innovative approach for automated fault localization utilizing a learning-to-rank  machine learning framework~\cite{joachims2002optimizing}. 
Learning-to-rank has been 
adopted by existing work for 
both IR-based~\cite{ye2014learning} and spectrum-based~\cite{b2016learning} fault localization.
The key idea of using Learning-to-Rank for fault localization is to train a machine learning model to rank pieces of software code by how likely they are to contain a bug, based on their relevance to a given bug report.
For example, Ye et al.~\cite{ye2014learning} employed a Learning-to-Rank approach to prioritize relevant files in bug reports through the integration of domain knowledge. 

\Name{} defines its ranking function as a weighted combination of features that comprehensively capture bug report semantic meanings, distinguishing it from existing techniques. Specifically, \Name{} harnesses the state-of-the-art Large Language Model (LLM), GPT-4, for this purpose. Moreover, \Name{} proposes new features tailored for learning-to-rank models based on the newly constructed queries by LLMs. Additionally, we present \Name{}+ as an enhanced version integrating GPT-4's conversational capabilities for iterative query reformulation based on user feedback, enhancing query accuracy. Evaluation against seven contemporary IRFL techniques using a 6,340-bug report dataset demonstrates \Name{}+'s superior performance. With a high Mean Reciprocal Rank (MRR) of 0.6770 and Mean Average Precision (MAP) of 0.5118, \Name{}+ outperforms all seven state-of-the-art IRFL techniques. Notably, \Name{} pioneers LLM application to query construction in IRFL.

In summary, this paper makes the following contributions:

\begin{itemize}
    \item A novel query construction approach for IRFL using Large Language Models (LLMs).
    \item Development of a user feedback-driven query reformulation process, including a bias-mitigation template, improving query precision and reliability.
    \item Strategic use of key features and a learning-to-rank model to enhance IRFL, ensuring accurate fault localization.
    \item Provision of a tool and experimental dataset publicly available\footnote{https://github.com/jedishao/LLmiRQ}.
\end{itemize}

\section{Background and Motivation }
\label{back}

In this section, we describe the background and provide motivating examples.

\subsection{Background}

Information Retrieval (IR)-based fault localization is a technique that utilizes information retrieval methods to pinpoint potential bug locations in software. This approach capitalizes on the textual information surrounding software elements, such as source code, comments, and error logs, as it often contains crucial insights into the functionality and potential faults of these elements. By treating source code and bug reports as textual documents, IR-based fault localization employs text similarity measures to align the descriptions in bug reports with the corresponding parts of the source code.

The pivotal components of IRFL encompass the query and retrieval models, where significant research has been dedicated to enhancing and refining both aspects. 
%
For improving query construction, Saha et al.\cite{bluir} utilized the Eclipse JDT to effectively process the source code's Abstract Syntax Tree (AST), extracting vital information from various code entities. This significantly enhances the efficiency of IR-based fault localization by constructing more precise queries. Rahman et al.~\cite{blizzard} incorporates context-aware query reformulation into bug localization, using a graph-based approach to represent bug reports and employing PageRank to identify important tokens for queries, demonstrating some improvements. However, as highlighted in subsequent sections, existing queries often 
\emph{lack the accuracy or comprehensiveness} necessary for effective fault localization.

Efforts have also been made to enhance retrieval models in fault localization~\cite{blia,blizzard,bluir,amalgam}. For example, Wang et al.\cite{amalgam} proposed a method that leverages version history information to refine IRFL techniques, thereby improving the model’s ability to adeptly exploit historical data. Beyond static retrieval models, learning-based models are also utilized, such as Learning-to-Rank (LtR)~\cite{joachims2002optimizing}. LtR represents a machine learning methodology used to develop ranking models for information retrieval systems. The LtR process begins with feature extraction from bug reports and source code, including textual similarity scores and code metrics. By leveraging historical bug reports and their corresponding fixes, training data is constructed. This data comprises queries (bug reports), candidate documents (source code files or methods), and relevance labels indicating the presence of faults. This data is used to train the LtR model to rank candidate documents based on their likelihood of containing faults.

Several methodologies have surfaced that harness Learning-to-Rank (LtR) for fault localization from bug reports. For instance, Ye et al.~\cite{ye2014learning} utilized a Learning-to-Rank strategy to prioritize pertinent files in bug reports by integrating domain knowledge. Likewise, Le et al. \cite{b2016learning} applied Learning-to-Rank techniques to bolster spectrum-based fault localization methodologies.
Despite these advancements, current approaches grapple with suboptimal accuracy primarily due to 
\emph{imprecise feature representation} that fails to accurately capture the mapping between bug reports and source code.

\subsection{Motivation}
\label{background}



\begin{table}[]\scriptsize
\setlength{\tabcolsep}{2pt} 
\caption{Motivation Examples of Bug Reports and Queries.}\label{moti}
\begin{tabular}{p{\linewidth}} 
\hline
\multicolumn{1}{c}{\textbf{COMPRESS-357}} \\
\hline
\textbf{Title}: BZip2CompressorOutputStream can affect output stream incorrectly \\
\textbf{Description}: BZip2CompressorOutputStream has an unsynchronized finished() method, and an unsynchronized finalize method. Finish checks to see if the output stream is null, and if it is not, it calls various methods, some of which write to the output stream. Now, consider something like this sequence. BZip2OutputStream s = ... s.close(); s = null; After the s = null, the stream is garbage. At some point the garbage collector call finalize(), which calls finish(). \ldots \\ \hline
\textbf{BugLocator}: title + description \textbf{Rank}: 11 \\
\hline
\textbf{BLIZZARD}: finish stream calls output method class synchronize time bzip compressor compressoroutputstream output stream compressor stream affect incorrectly \textbf{Rank}: 47 \\
\hline
\textbf{GPT-4}: BZip2CompressorOutputStream finalize finish \textbf{Rank}: 1 \\
\hline
\multicolumn{1}{c} {\textbf{CAMEL-620}} \\\hline
\textbf{Title}: ResequencerType.createProcessor could throw NPE as stream config does not get initialized.\\
\textbf{Description}: java.lang.NullPointerException\\
	at org.apache.camel.model.ResequencerType.createProcessor(ResequencerType.java:163)\\
	at org.apache.camel.model.ProcessorType.createOutputsProcessor(ProcessorType.java:584)\\
	at org.apache.camel.model.ProcessorType.createOutputsProcessor(ProcessorType.java:93)\\
        \dots \\\hline
\textbf{BugLocator}: title + description  \textbf{Rank}: 223 \\ \hline
\textbf{BLIZZARD}:  NullPointerException createStreamResequencer createProcessor create  
  createOutputsProcessor ResequencerType addRoutes ProcessorType Processor RouteType 
 startRouteDefinitions   InterceptorRef   \textbf{Rank}: 2659  \\\hline
\textbf{GPT-4}: ResequencerType ResequencerTest createProcessor  \textbf{Rank}: 2 \\\hline
\multicolumn{1}{c} {\textbf{CAMEL-2320}}\\\hline
\textbf{Title}: JDBC component doesn't preserve headers\\
\textbf{Description}: JDBC component doesn't preserve any of the headers that are sent into it\\\hline
 \textbf{BugLocator}: title + description  \textbf{Rank}: 179 \\ \hline
 \textbf{BLIZZARD}: set jdbc ftp row parameters generated endpoint output JDBC component 
 preserve headers JDBC component  preserve headers  \textbf{Rank}: 7  \\\hline
\textbf{GPT-4}: JdbcProducer JdbcEndpoint JdbcComponent  \textbf{Rank}: 3 \\\hline
\end{tabular}
\vspace*{-10pt}
\end{table}

\subsubsection{Query Construction} 

Table~\ref{moti} demonstrates the limitations of current methods and the potential improvements using GPT-4 for different types of bug reports.

In COMPRESS-357, traditional methods like BugLocator and BLIZZARD struggle with noisy data. GPT-4, however, understands the bug report and pinpoints the root cause by identifying crucial class names like 'BZip2CompressorOutputStream', leading to more accurate rankings.
For CAMEL-620, which includes stack traces, existing tools often fail to analyze them effectively. 
This is because they primarily focus on the position, such as the top 10 entries of stack traces, without truly understanding the context or the content of the stack trace itself.
GPT-4 can analyze the stack trace and identify the root cause by pointing to specific classes and methods, enhancing fault localization accuracy.
In CAMEL-2320, the bug report contains only natural language with little useful information. Traditional methods fall short here, but GPT-4 can extend the information by understanding the context and identifying relevant classes and methods, thereby improving the accuracy of fault localization.

\subsubsection{Retrieval Model}

Furthermore, for bug reports CAMEL-620 and CAMEL-2320 in Table~\ref{moti}, the root causes pinpointed by GPT-4—ResequencerType and JdbcProducer, respectively—do not achieve the top rank position when using conventional retrieval models like BugLocator's rVSM. The challenge lies in the nature of queries generated by GPT-4, which consist of programming entities. Traditional retrieval models that calculate similarity between these queries and the source code often fail to filter out noise present in the source code, such as comments or irrelevant programming entities. This observation leads to the hypothesis that a \emph{more precise} approach, focusing on exact matches of class and method names, could yield improvements.

Additionally, empirical investigations suggest a tendency for files implicated in a single bug report to exhibit call relationships~\cite{li2021laprob}.
For instance, in the case of CAMEL-620, not only is \texttt{ResequencerType} identified as the buggy file, but associated files like \texttt{ResequencerTest} and \texttt{SteamResequencerType}, which interact with \texttt{ResequencerType}, also require attention. This interconnection underlines the potential benefits of considering call relationships in enhancing retrieval models.

In summary, traditional retrieval models are not optimally aligned with the unique characteristics of queries generated by advanced language models like GPT-4. To bridge this gap, we propose an \emph{enriched feature set} for a learning-to-rank model, encompassing class name match score, call graph score, and five features adopted from existing work~\cite{ye2014learning}.
Integrating these features into a learning-to-rank framework is expected to significantly augment the efficacy of IR-based fault localization.
\section{Approach}

\begin{figure*}
    \centering
    \includegraphics[width=1\textwidth, height=0.38\textheight]{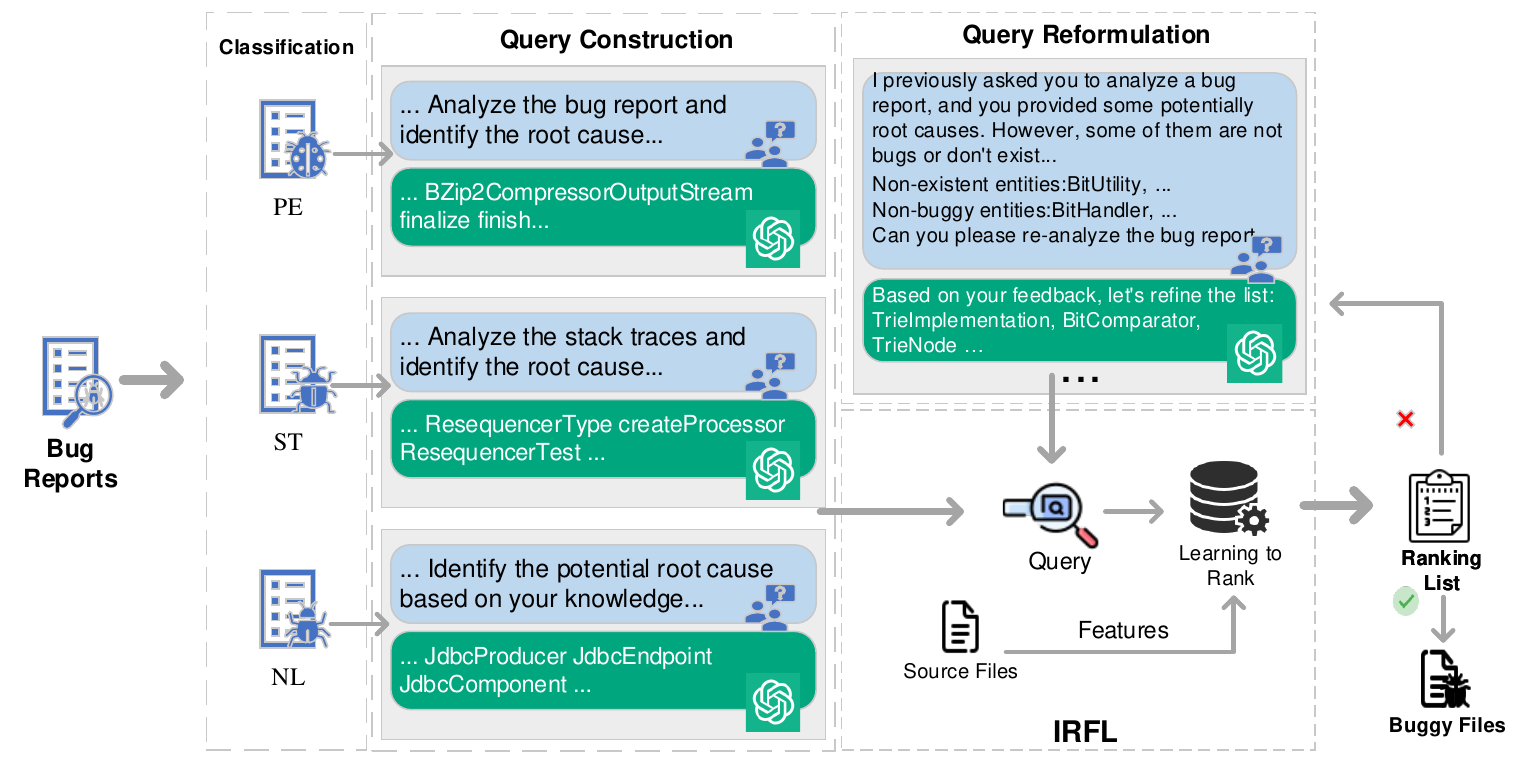}
    \caption{The overview of \Name{} and \Name{}+.} \label{overview}
    \vspace*{-10pt}
\end{figure*}

Figure~\ref{overview} provides an overview of our approach. Given that different bug report components, 
such as programming entities and stack traces, can have significantly varying impacts on IRFL, 
tailored approaches are required for analysis. 
To begin, we segment bug reports into three distinct categories based on their content. We first determine if a report contains a stack trace; if it does, we categorize it as ST (Stack Trace). If not, we then check for the presence of programming entities; if present, the report is categorized as PE (Programming Entities). Reports that contain neither stack traces nor programming entities are classified as NL (Natural Language).

The rationale behind the categories is that for PE and ST bug reports, which often contain precise details such as specific class or method names, a \emph{query reduction} strategy is applied to minimize noise and highlight the critical information. In contrast, NL reports usually lack direct code references, necessitating a \emph{query extension} approach to incorporate broader contextual information for effective fault localization.
This approach is referred to as \Name{}.

Each category is subjected to a customized query construction strategy, which serves as the initial query for IRFL.
Our approach uses iterative design of prompts for each category to optimize the query construction process:
Prompts are initially designed based on the type of information present in the bug report.
Prompts are evaluated and refined based on feedback and empirical results to improve their effectiveness.
The refined prompts are used to generate precise queries tailored to the content of each bug report category. The detail of the 
prompts are discussed in Section 3.2.

Upon generating the initial IRFL query, we obtain a results list. In cases where the initial IRFL results fail to precisely identify the problematic files, we introduce a \emph{user and conversation-driven query reformulation} process facilitated by GPT. This enhanced approach is denoted as \Name{}+. Through this iterative refinement, we aim to achieve more accurate and effective fault localization.

Besides query construction, 
our approach enhances IR-based fault localization by leveraging a learning-to-rank (LtR) model, which is trained using historical bug reports and their corresponding fixes. 
The LtR model ranks the source files, providing users with a list where higher-ranked files are more likely to contain the bug.
We design two novel features to capture various aspects of the bug reports and source code: the \emph{class name match score} and the \emph{call graph score}. The class name match score helps identify relevant classes involved in the bug, while the call graph score captures dynamic relationships between code components. Additionally, we incorporate five features from existing LtR works, such as text similarity, historical bug fix frequency, and recency, to enhance our model’s effectiveness.


\subsection{Bug Reports Classification} 
\label{classify}

As described in Section~\ref{back}, the quality of bug reports plays an important role in IR-based fault localization. 
Different kinds of bug reports need different query construction strategies. 

\subsubsection{Text Containing Programming Entities (PE)}
Program elements include method names, package names, and source file names. Bug reports containing one or more of these elements, but no stack traces, are classified as \textbf{PE}. Queries from these reports are rich in information. Appropriate regular expressions~\cite{rigby2013discovering} are applied to identify these elements within the text.

\subsubsection{Stack Traces (ST)} 
Stack traces include sequences of method calls during an error, often pinpointing the exact class or method where the error occurred. Bug reports with one or more stack traces are labeled as \textbf{ST}. Since stack traces contain structured information, queries generated can be noisy. We use specific regular expressions~\cite{moreno2014use} to identify relevant trace entries.

\subsubsection{Pure Text (NL)} 

Bug reports described solely in plain natural language, without specific references to programming entities or stack traces, are classified as \textbf{NL}. These reports pose challenges due to the lack of direct code-related terms, requiring reliance on the textual content's accuracy.

We build a classifier by leveraging previous work~\cite{blizzard} to automatically categorize bug reports. Using this classifier, \Name{} flags a given bug report into a specific category and applies the corresponding strategy to construct the query (Figure 1).

\subsection{Query Construction}

By leveraging GPT's ability to comprehend and contextualize diverse textual information, our goal is to explore its potential in assisting developers in efficiently identifying bugs. This section outlines our three core query construction strategies: query reduction, query expansion, and query reformulation, and describes their utilization in IR-based fault localization.

\subsubsection{Query Reduction}
Query reduction aims to highlight the most important tokens and discard superfluous ones, a task where traditional methods often fall short. GPT-4's advanced contextual understanding enhances the precision of identifying root causes, thus revolutionizing query reduction.
The prompt used is: 

\textit{“Analyze the bug report and construct a query by identifying programming entities (e.g., classes, methods) that may be relevant to the bug's root cause.”}

This prompt was iteratively designed through the following steps:
\textit{Initial Query Construction}: The initial prompt focused broadly on identifying key elements in the bug report. Early versions did not specify the type of programming entities, which resulted in less targeted queries.
\textit{Feedback and Refinement}: Feedback indicated the need for more precise identification of programming entities. As a result, the prompt was refined to specify "classes" and "methods" explicitly, directing the model's attention to these crucial elements.
\textit{Empirical Testing}: The refined prompt was tested on various bug reports to ensure it consistently produced accurate and relevant queries.
For example, in bug report COMPRESS-357, GPT-4 generates the query: "BZip2CompressorOutputStream finalize finish," derived from the analysis of the bug report.

For stack traces (ST), which inherently carry crucial information, GPT-4 can accurately identify relevant classes and methods. 
The prompt used is:

\textit{“Analyze the provided stack traces and construct a query, identifying programming entities (e.g., classes, methods) relevant to the bug's root cause.”}

This prompt was iteratively designed as follows:
Initial Query Construction: The initial prompt was general and did not focus specifically on stack traces. It was found to be less effective in extracting relevant information from stack traces.
Feedback and Refinement: The prompt was adjusted to explicitly mention "stack traces" and to identify "classes" and "methods," which are often the key elements in stack traces.
Empirical Testing: The refined prompt was tested on bug reports with stack traces, demonstrating improved accuracy in identifying relevant programming entities.

In bug report CAMEL-620, the query generated includes entities like "ResequencerType ResequencerTest createProcessor."

\subsubsection{Query Expansion}
Query expansion is essential for bug reports that lack informative details (NL), such as specific programming entities or stack traces. Traditional methods require significant human effort and often do not consistently improve IRFL outcomes. GPT-4, with its vast internal knowledge, can introduce relevant programming entities, enriching the available information.
The prompt used is:

\textit{“Analyze the bug report and construct a query by identifying potential programming entities (e.g., classes, methods) relevant to the bug's root cause based on your knowledge.”}

This prompt was iteratively designed through the following steps:
Initial Query Construction: The initial prompt was very broad and did not effectively guide the model to identify potential programming entities.
Feedback and Refinement: Feedback highlighted the need for the model to draw upon its extensive internal knowledge. The prompt was refined to specify "potential programming entities (e.g., classes, methods)" to guide the model more effectively.
Empirical Testing: The refined prompt was tested on bug reports lacking specific details, showing improved performance in generating insightful queries.

After expanding the query, a straightforward codebase analysis is conducted to ensure the suggested classes or methods exist within the project. Non-existent entities are removed.
For example, in bug report CAMEL-2320, GPT-4 might suggest entities like "JdbcProducer," "JdbcEndpoint," and "JdbcComponent," inferred from the context of the bug report.

By adopting these tailored strategies and iteratively designing prompts, our approach enhances the overall performance of IRFL, addressing the limitations of traditional models and making full use of the precise queries generated by GPT-4.

\subsection{Interactive Query Reformulation}

When the initial query does not produce satisfactory results, reformulating the query could help optimize the search process. 
Inspired by existing works~\cite{chaparro2019using, florez2021combining, xia2023keep}, 
we have designed a user and conversational-driven query reformulation using GPT. 
The envisioned usage scenario begins with the execution of an initial query. 
Subsequently, users examine the results list (the top 10 files) and, 
if the buggy files are not present in this list, they can provide feedback to GPT. 

To minimize bias in the recommendations provided by LLMs while maintaining the feedback's informativeness and value, we've designed two types of feedback mechanisms. 
The first is \textbf{Feedback on Non-Existing Classes}, where users report classes suggested by the AI that don't exist in the source code. 
For example, consider a situation like the query reformulation depicted in Figure~\ref{overview}. 
In this scenario, a user provides feedback stating, "I couldn’t find any class named BitUtility in the source code." 
Such feedback assists the model in refining its suggestions by excluding non-existing or irrelevant classes, thereby improving the precision of its recommendations. 
The second type of feedback is \textbf{Feedback on Non-Buggy Classes}. 
In this case, users inform the model about classes that were suggested but are not related to the bug (non-buggy). 
For instance, refer to the second feedback provided in Figure~\ref{overview}, where the user states, "TrieImplementation doesn't seem to have the issue." 
The model utilizes this feedback to narrow down its suggestions, concentrating on classes that are more likely to be buggy. 
This approach enhances the relevance and utility of the model's recommendations.

Taking into account both user experience and computational efficiency, we've established a maximum of 5 conversation iterations. 
Persistently soliciting feedback without yielding satisfactory results can prove to be both time-consuming and exasperating for users. 
If the buggy files remain elusive even after 5 query reformulations based on user insights, 
it may suggest inherent shortcomings in the query formulation methodology or inconsistencies in the underlying data. In such cases, the query reformulation is deemed unsuccessful. 
Conversely, if a reformulated query successfully pinpoints the buggy files within the result list, then the query reformulation is considered a success.

\subsection{Feature Selection}
\label{fs}

Common Information Retrieval-based Fault Localization (IRFL) techniques often utilize bug reports as queries. 
These reports typically include extraneous information that introduces noise into the retrieval process. 
In contrast, our method constructs queries composed exclusively of programming entities. 
This focused approach necessitates a departure from traditional text similarity measures, 
which may also amplify noise due to their inclusion of less relevant textual data in source code. 
To harness the full potential of our refined queries, 
we have carefully curated a set of features designed to capitalize on the precise nature of the programming entities identified by GPT-4. 
These features aim to enhance the alignment between query terms and relevant source code, 
thereby improving the accuracy and efficiency of the fault localization process. 

\subsubsection{Class Name Match Score}

Given that our queries are comprised solely of programming entities, 
specifically class and method names, these entities offer a direct and potent indication of potentially buggy files. 
Our first selected feature, the Class Name Match Score, is designed to exploit this correlation. 
For each source file \(s\) that contains a class name appearing in our query \(q\), 
we calculate this feature based on the length of the class name in question (\(s.class\)). 

We opt for a score proportional to the name's length, 
rather than a binary presence indicator (1 or 0), to facilitate a more nuanced ranking in the fault localization process. 
This decision stems from the premise that longer class names tend to be more specific. 
Consequently, a bug report that explicitly mentions a lengthy and specific class name is likely pointing directly towards the buggy file. 
Therefore, by evaluating files based on the length of the matched class name, 
we aim to prioritize files with more specific (and thus potentially more relevant) class names in the ranking list.

\[
f_1(s, q) = |s.class| \quad \text{if } s.class \in q
\]

This approach ensures that the ranking not only highlights files with matching class names 
but also subtly distinguishes them based on the specificity implied by the length of these names, 
thereby enhancing the effectiveness of the fault localization.

\subsubsection{Call Graph Score}

The textual similarity between source files, though insightful, may not fully capture the complexity of their interdependencies. 
Our empirical investigations support the notion that source files associated with the same bug often share call relationships. 
This observation aligns with our initial premise stemming from the Class Name Match Score (Feature 1), 
suggesting that programming entities referenced in bug reports are likely to pinpoint the locus of bugs. 
It follows, then, that the scope of these entities should extend beyond individual files to encompass related files through call relationships.

In pursuit of this extended scope, we propagate the Class Name Match Score throughout the call graph. 
The Call Graph Score \( f_2 \) for a file \( s_k \) is thus calculated by aggregating the scores from its interconnected files:

\[ f_{2}(s_k) = \sum_{s \in \text{Callers}(s_k)} f_{1}(s) + \sum_{s \in \text{Callees}(s_k)} f_{1}(s) \]

Here, \( \text{Callers}(s) \) and \( \text{Callees}(s) \) represent the sets of files that call and are called by \( s \), respectively. 
The Class Name Match Score, \( f_1(s) \), quantifies the relevance of each file \( s \) based on the mentioned class names.

This formula encapsulates the aggregated impact of both upstream and downstream dependencies, 
reflecting the compounding risk of bugs disseminated through the call graph. 
It thus offers a nuanced prediction of a file’s susceptibility to bugs based on its contextual role within the software architecture.

Employing the Call Graph Score enables a multidimensional analysis that fuses direct textual correlations with the structural dynamics of software interactions. 
The result is a holistic and nuanced assessment of each file's potential faultiness, enhancing the precision of the fault localization process.

\subsubsection{Features from Existing Work}

We incorporate five features from previous studies~\cite{ye2014learning} that also utilize learning-to-rank for IRFL.

\textit{Text Similarity Score:}
The similarity between a bug report \( r \) and a source file \( s \) is quantified using the cosine similarity measure, 
defined as the cosine of the angle between their respective tf-idf vectors.
%
%
This metric is particularly adept at highlighting textual relationships, ranging from explicit term matches to more subtle linguistic correlations.

\textit{API-Enriched Lexical Similarity:} This feature measures the similarity between a bug report and the technical terms found in source code APIs. It is defined as:

$$f_{4}(r, s) = \max\{\text{sim}(r, s.\text{api}), \text{sim}(r, m.\text{api}) \mid m \in s\}$$

where \( \text{sim} \) is the similarity measure between the bug report \( r \) and the API documentation for the source file \( s \) or its methods \( m \).

\textit{Collaborative Filtering Score:} This score evaluates the similarity between a bug report and historical reports of the same source file, leveraging past bug fix data:

\[ f_{5}(r, s) = \text{sim}(r, \text{br}(r, s)) \]

\textit{Bug-Fixing Recency:} This feature reflects the time since the last bug fix, calculated as:

\[ f_{6}(r, s) = \frac{1}{(r.\text{month} - \text{last}(r, s).\text{month} + 1)} \]

More recent fixes yield a higher score.

\textit{Bug-Fixing Frequency:} This metric counts the number of previous fixes for a file:

\[ f_{7}(r, s) = |\text{br}(r, s)| \]

Files with more past fixes receive a higher score.














\subsection{Feature Normalization}

In preparation for training the ranking models, it is crucial to normalize the feature values to a consistent range, 
specifically [0, 1]. This normalization is conducted according to the following scheme:
$$
f^{'}_{i}=\left\{\begin{matrix}
0 & \text{if} \ f_{i}<f_{i}.min\\ 
\frac{f_{i}-f_{i}.min}{f_{i}.max-f_{i}.min} & \text{if} \ f_{i}.min\leqslant f_{i}\leqslant f_{i}.max \\ 
1 & \text{if} \ f_{i} > f_{i}.max
\end{matrix}\right. 
$$

Here, \( f_i \) represents the original value, while \( f'_i \) denotes the normalized value of the \( i^{th} \) 
feature associated with a potentially buggy method. The terms \( \text{min}_i \) and \( \text{max}_i \) are the minimum and maximum values of the \( i^{th} \) feature, as determined from the training dataset.

\subsection{Learning-to-Rank for Fault Localization} 

Learning-to-Rank (LtR) models have revolutionized the field of information retrieval by providing a mechanism to learn complex ranking functions from data. 
In the domain of fault localization, an LtR model can be trained to rank source files in order of their likelihood of containing bugs, as indicated by a given bug report. 

SVM$^{rank}$ is a specialized instance of Support Vector Machines (SVMs) tailored for ranking tasks, as proposed by T. Joachims. 
It operates by learning a linear ranking function that minimizes the average number of misordered pairs of documents (or in our case, source files).
SVM$^{rank}$ is chosen for its efficiency in handling sparse feature spaces and its effectiveness in learning with linear kernels, 
which is particularly beneficial given the high dimensionality typical of text data in software repositories. 
Additionally, SVM$^{rank}$'s capacity to process large datasets makes it well-suited for the vast amounts of source code and bug report data.

\subsubsection{Training Process}

These features are combined in a single feature vector to train the SVM$^{rank}$ model, 
allowing for a comprehensive and nuanced ranking of source files according to their likelihood of containing bugs.
To train our bug report classification model, we utilized a chronological cross-validation approach. 
Since different bug reports types have different characteristics, 
this involved dividing our dataset into 10 subsets for each categories (i.e., PE, ST, NL) based on chronological order, 
ranging from the oldest (\(Subset_{1}\)) to the most recent (\(Subset_{10}\)). 
We trained our model on earlier subsets and tested it on later subsets, 
mimicking real-world scenarios where models are trained on recent data and tested on future data. 
This process ensures that our bug report classification model is robust and generalizable, 
with evaluations conducted using standard metrics to assess performance and iteratively improve the model's accuracy and reliability.

\subsubsection{Fault Localization}

Our fault localization process involves giving a bug report to GPT-4, 
which will analyze it and generate queries based on our prompts. We then extract each feature for our trained learning-to-rank model. 
The model computes a score for each source file in the software project and uses this value to rank all the files in descending order. 
Users receive a ranked list of files, where higher-ranked files are more likely to be relevant to the bug report, 
indicating a higher likelihood of containing the bug's cause.

\section{Evaluation}
\label{evaluation}

We evaluate the effectiveness of queries generated by GPT-4 using various prompts to address the following research questions: 

\vspace*{3pt}
\noindent
\textbf{RQ1: How does the performance of \Name{} in IRFL compare to the state-of-the-art techniques?}
The goal of this research question is to assess the effectiveness of \Name{} in fault localization when compared to the state-of-the-art techniques. 

\vspace*{3pt}
\noindent
\textbf{RQ2: How does the effectiveness of our LtR model compare to state-of-the-art IR models?}
The goal of this research question is to evaluate the effectiveness of each feature and \Name{}'s performance across different retrieval models.

\vspace*{3pt}
\noindent
\textbf{RQ3: How effective are our queries in IRFL?}
The goal of this research question is to evaluate the effectiveness of our query compared to existing query construction methods, the number of conversational-based reformulations, and the differences between 0-shot and 1-shot scenarios.

\subsection{Experiment Setup}

\subsubsection{Dataset Collection} 
The dataset was collected by Lee et al.~\cite{bench4bl} and comprises 46 projects from Apache, Commons, JBoss, Spring, and Wildfly, with a total of 9,459 bug reports. The dataset is employed in a reproducibility study focused on the performance of IRFL. The size of this dataset is comparable to those utilized in existing IRFL studies~\cite{buglocator, brtracer, blia, amalgam, bluir, blizzard}, which range from 3,479 to 5,139 bug reports.
Typically, Users submit bug reports in relation to specific versions of a project. 
Regarding standard IRFL evaluation, a common approach involves a single version matching strategy, 
assuming the search encompasses potential source code files from a project's most recent version. 
Our evaluation also adopts a single version matching strategy. 
However, due to the inclusion of some bug reports originating from very old project versions, corresponding source files are often unavailable. 
The absence of these source files introduces potential ambiguity and uncertainty into our evaluation process. 
As a solution, we opt to eliminate bug reports lacking corresponding fixed source files from the dataset.  
This process aims to ensure the quality and reliability of the dataset, leading to a more accurate representation of the data and enhancing the validity of our research results.
After the process, the dataset comprises a total of 6,340 bug reports, as shown in Table~\ref{dataset}. 

\begin{table}[]\scriptsize
\setlength{\tabcolsep}{2pt} 
\caption{Dataset.}\label{dataset}
\begin{tabular}{|l|c|c|c|c|l|c|c|c|c|}
\hline
\textbf{Projects} & \textbf{\#PE} & \textbf{\#ST} & \textbf{\#NL} & \textbf{\#BR} & \textbf{Projects} & \textbf{\#PE} & \textbf{\#ST} & \textbf{\#NL} & \textbf{\#BR} \\ \hline \hline
CAMEL             & 603           & 113           & 206           & 922           & DATAREDIS         & 39            & 8             & 0             & 47            \\ \hline
HBASE             & 334           & 117           & 59            & 510           & DATAREST          & 77            & 16            & 15            & 108           \\ \hline
HIVE              & 467           & 190           & 381           & 1038          & LDAP              & 38            & 7             & 3             & 48            \\ \hline
CODEC             & 38            & 2             & 1             & 41            & MOBILE            & 11            & 0             & 0             & 11            \\ \hline
COLLECTIONS       & 80            & 3             & 1             & 84            & ROO               & 100           & 30            & 63            & 193           \\ \hline
COMPRESS          & 84            & 17            & 8             & 109           & SEC               & 259           & 42            & 20            & 321           \\ \hline
CONFIGURATION     & 87            & 14            & 2             & 103           & SECOAUTH          & 34            & 5             & 14            & 53            \\ \hline
CRYPTO            & 1             & 0             & 0             & 1             & SGF               & 52            & 17            & 10            & 79            \\ \hline
CSV               & 12            & 1             & 1             & 14            & SHDP              & 27            & 5             & 12            & 44            \\ \hline
IO                & 71            & 8             & 2             & 81            & SHL               & 2             & 0             & 6             & 8             \\ \hline
LANG              & 147           & 15            & 1             & 163           & SOCIAL            & 12            & 1             & 0             & 13            \\ \hline
MATH              & 190           & 9             & 6             & 205           & SOCIALFB          & 11            & 3             & 1             & 15            \\ \hline
WEAVER            & 0             & 0             & 2             & 2             & SOCIALLI          & 4             & 0             & 0             & 4             \\ \hline
ENTESB            & 0             & 2             & 2             & 4             & SOCIALTW          & 8             & 0             & 0             & 8             \\ \hline
JBMETA            & 8             & 10             & 1             & 10            & SPR               & 74            & 22            & 7             & 103           \\ \hline
AMQP              & 59            & 18            & 13            & 90            & SWF               & 69            & 2             & 11            & 82            \\ \hline
ANDROID           & 4             & 2             & 2             & 8             & SWS               & 79            & 25            & 6             & 110           \\ \hline
BATCH             & 247           & 23            & 28            & 298           & ELY               & 17            & 0             & 4             & 21            \\ \hline
BATCHADM          & 11            & 2             & 5             & 18            & SWARM             & 29            & 9             & 7             & 45            \\ \hline
DATACMNS          & 100           & 28            & 8             & 136           & WFARQ             & 1             & 0             & 0             & 1             \\ \hline
DATAGRAPH         & 2             & 0             & 0             & 2             & WFCORE            & 156           & 68            & 100           & 324           \\ \hline
DATAJPA           & 99            & 27            & 10            & 136           & WFLY              & 238           & 133           & 115           & 486           \\ \hline
DATAMONGO         & 165           & 58            & 16            & 239           & WFMP              & 2             & 0             & 0             & 2             \\ \hline
                  &               &               &               &               & \textbf{Total}                  & 4148          & 1043          & 1149          & 6340          \\ \hline
\end{tabular}
\vspace*{-15pt}
\end{table} 

\subsubsection{Bug Reports Classification} 
As outlined in Section 3, the quality of bug reports is vital for IRFL effectiveness. Table~\ref{dataset} categorizes bug reports based on their content: \textbf{PE} denotes reports with programming entities in their descriptions. \textbf{ST} refers to reports containing stack traces. \textbf{NL} represents reports composed solely of plain natural language, excluding programming entities and stack traces. 
\textbf{BR} refers to the total number of bug reports.
These categories are mutually exclusive; for instance, a report featuring stack traces is classified under ST regardless of the presence or absence of programming entities in its description.


\subsubsection{IRFL Techniques}

We select seven IRFL techniques to assess the effectiveness of \Name{} and \Name{}+ in IRFL tasks. 
Table~\ref{irbltech} presents an overview of the settings for each technique. 
Each of these techniques considers various resources, which are generally classified into two categories: Bug Report and External Features. 
Bug Report resources encompass text descriptions (BRT) and stack traces (ST), 
while External Features resources encompass version history (VH), similar bug reports (SB), and source files (SF). 
Specifically, \textbf{BugLocator}~\cite{buglocator} suggests relevant files by examining similar bugs through rVSM. 
\textbf{BLUiR}~\cite{bluir} differentiates between summary and description in bug reports, employing distinct query and document representations for separate searches.
\textbf{BRTracer}~\cite{brtracer} extends BugLocator by incorporating segmentation and stack-trace analysis to rank files. 
\textbf{AmaLgam}~\cite{amalgam} employs three score components (version history, similar report, and structure) to rank files. 
\textbf{BLIA}~\cite{blia} adapts the approaches of BugLocator, BLUiR, and BRTracer, employing Google’s algorithm for version history analysis.
\textbf{BLIZZARD}~\cite{blizzard} transforms bug report text, stack traces, and source code into graphs, utilizing graph-based term weighting and Lucene for fault localization. 
\textbf{LR}~\cite{ye2014learning} uses text similarity, API specification similarity, bug fix frequency, and bug fix recency as features, employing learning-to-rank (LtR) for training.

\begin{table}[]\footnotesize
    \centering
    \caption{Comparison of IRFL Techniques.}\label{irbltech}
    \begin{tabular}{lcccccc}
    \hline
    \multicolumn{1}{|l|}{\multirow{2}{*}{Technique}} & \multicolumn{2}{c|}{Bug Report}                                                   & \multicolumn{3}{c|}{External Features}                                                     & \multicolumn{1}{c|}{\multirow{2}{*}{IR Model}} \\ \cline{2-6}
    \multicolumn{1}{|l|}{}                           & \multicolumn{1}{c|}{\textbf{BRT}}       & \multicolumn{1}{c|}{\textbf{ST}}        & \multicolumn{1}{c|}{\textbf{VH}}        & \multicolumn{1}{c|}{\textbf{\textbf{SB}}}        & \multicolumn{1}{c|}{\textbf{SF}}  & \multicolumn{1}{c|}{} \\ \hline
    \multicolumn{1}{|l|}{BugLocator}                 & \multicolumn{1}{c|}{$\bullet$} & \multicolumn{1}{c|}{}          & \multicolumn{1}{c|}{}          & \multicolumn{1}{c|}{$\bullet$} & \multicolumn{1}{c|}{}          & \multicolumn{1}{c|}{rVSM}                      \\ \hline
    \multicolumn{1}{|l|}{BLUiR}                      & \multicolumn{1}{c|}{$\bullet$} & \multicolumn{1}{c|}{}          & \multicolumn{1}{c|}{}          & \multicolumn{1}{c|}{$\bullet$} & \multicolumn{1}{c|}{$\bullet$} & \multicolumn{1}{c|}{Indri}                     \\ \hline
    \multicolumn{1}{|l|}{BRTracer}                   & \multicolumn{1}{c|}{$\bullet$} & \multicolumn{1}{c|}{$\bullet$} & \multicolumn{1}{c|}{}          & \multicolumn{1}{c|}{$\bullet$} & \multicolumn{1}{c|}{$\bullet$} & \multicolumn{1}{c|}{rVSM}                      \\ \hline
    \multicolumn{1}{|l|}{AmaLgam}                    & \multicolumn{1}{c|}{$\bullet$} & \multicolumn{1}{c|}{}          & \multicolumn{1}{c|}{$\bullet$} & \multicolumn{1}{c|}{$\bullet$} & \multicolumn{1}{c|}{$\bullet$} & \multicolumn{1}{c|}{Mixed}                     \\ \hline
    \multicolumn{1}{|l|}{BLIA}                       & \multicolumn{1}{c|}{$\bullet$} & \multicolumn{1}{c|}{$\bullet$} & \multicolumn{1}{c|}{$\bullet$} & \multicolumn{1}{c|}{$\bullet$} & \multicolumn{1}{c|}{$\bullet$} & \multicolumn{1}{c|}{rVSM}                      \\ \hline
    \multicolumn{1}{|l|}{BLIZZARD}                   & \multicolumn{1}{c|}{$\bullet$} & \multicolumn{1}{c|}{$\bullet$} & \multicolumn{1}{c|}{}          & \multicolumn{1}{c|}{}          & \multicolumn{1}{c|}{$\bullet$} & \multicolumn{1}{c|}{Lucene}                    \\ \hline
    \multicolumn{1}{|l|}{LR}                   & \multicolumn{1}{c|}{$\bullet$} & \multicolumn{1}{c|}{} & \multicolumn{1}{c|}{$\bullet$}          & \multicolumn{1}{c|}{$\bullet$}          & \multicolumn{1}{c|}{} & \multicolumn{1}{c|}{LtR}                    \\ \hline
    \multicolumn{7}{l}{\begin{tabular}[c]{@{}l@{}}   \footnotesize \textbf{BRT}: Bug Report Text, \textbf{ST}: Stack Trace, \textbf{VH}: Version History, \\ \footnotesize  \textbf{SF}: Source File, \textbf{SB}: Similar Bug Report\end{tabular}}                                                                             
    \end{tabular}
    \vspace*{-15pt}
    \end{table}

\subsubsection{Performance Metrics}

We use three performance metrics to evaluate our approach and other fault localization techniques. 
These metrics are widely used in the field and are also used by state-of-the-art techniques~\cite{buglocator, brtracer, blia, amalgam, bluir, blizzard, wen2016locus, ye2014learning}. 
    
\textbf{Top@K} measures the probability that the fault localization method will successfully locate the relevant bug reports when reviewing the first $k$ files (where $k=1, 5, 10$) in the recommendation list generated by the method. The calculation is $Top@K = \frac{|R_{k}|}{n} $, where
$|R_{k}|$ is the total number of bug reports that are successfully located when the method is recommended by Top$K$. $n$ is the total number of bug reports used in the evaluation process.
    
Mean Reciprocal Rank (\textbf{MRR})~\cite{MRR} measures the position of the first source file related to the bug report located by the fault localization method in the recommendation list. The calculation is
 $MRR = \frac{1}{n}\sum_{j=1}^{n}\frac{1}{rank_{j}}$, where $rank_{j}$ represents the ranking position of the first source files related to the $j$-th bug report in the recommendation list.

Mean Average Precision (\textbf{MAP})~\cite{MAP} measures the average position of all source files related to the bug report located by the fault localization method in the recommendation list. The calculation is 
    
$MAP = \frac{1}{n}\sum_{j=1}^{n}AvgP_{j}, \
AvgP_{j} = \frac{1}{|K_{j}|}\sum_{k\in K_{j}}^{}\{Prec@k\}$, 
$Prec@k = \frac{\sum_{i=1}^{k}IsRelevant(k)}{k} $, where $AvgP_{j}$ is the average precision for the $j$-th bug report, and $|K_{j}|$ is the total number of relevant source files for the $j$-th bug report. $Prec@k$ represents the precision of the top $k$ files in the recommendation list, and $IsRelevant(i)$ returns 1 if the $i$-th file in the recommendation list is related to the bug report, and 0 otherwise. $K_{j}$ is the true positive result set of a query.
The higher value of each metric represents better performance.

\subsection{Results and Analysis}
\label{ras}

\subsubsection{RQ1: How does the performance of \Name{} in IRFL compare to the state-of-the-art techniques?} 

To answer this RQ, we conducted an evaluation of \Name{} and \Name{}+ by comparing them with seven state-of-the-art 
IRFL techniques on our testing set, i.e., (\(Subset_{2}\)) to (\(Subset_{10}\)). 
To ensure fairness in the evaluation, we used the same configuration and parameters specified in each paper for all the techniques. 
Table~\ref{compare_stoa} presents the evaluation results of \Name{} and \Name{}+. 
The technique with the best performance on each metric is boldfaced. 
The result marked with "*" indicates the technique is not 
statistically distinguishable from the best algorithm at p = 0.05 using paired t-tests. 
Results without being boldfaced or "*" mean they are significantly lower than the best performance.
On average, \Name{} surpasses other methods across all five metrics, 
achieving a 0.6494 score on MRR and a 0.4841 score on MAP, 
besting the top-performing IRFL technique, BRTracer, by 9\% and 10\% respectively. 
Especially, compared to the existing learning-to-rank method LR, \Name{} outperformed it by 25\% on both MRR and MAP.

\begin{table}[t]\footnotesize
\setlength{\tabcolsep}{5pt}
\caption{Comparing with State-of-the-Art Techniques.} \label{compare_stoa}
\begin{tabular}{|c|l|c|c|c|c|c|}
\hline
                         & Tech       & Top1    & Top5    & Top10   & MRR    & MAP    \\ \hline \hline
\multirow{9}{*}{PE}      & BugLocator & 50.95\% & 75.92\% & 82.48\% & 0.6149 & 0.4657 \\ \cline{2-7} 
                         & BLUiR      & 42.19\% & 70.75\% & 78.68\% & 0.5402 & 0.4263 \\ \cline{2-7} 
                         & BRTracer   & 55.13\% & 77.87\% & 84.46\% & 0.6508 & 0.4875 \\ \cline{2-7} 
                         & AmaLgam    & 42.51\% & 71.04\% & 79.02\% & 0.5432 & 0.4293 \\ \cline{2-7} 
                         & BLIA       & 47.01\% & 74.90\% & 83.36\% & 0.5886 & 0.4632 \\ \cline{2-7} 
                         & BLIZZARD   & 40.08\% & 66.54\% & 74.90\% & 0.5128 & 0.3833 \\ \cline{2-7} 
                         & LR         & 48.38\% & 72.57\% & 81.01\% & 0.5860 & 0.4263 \\ \cline{2-7} 
                         & \Name{}     & \textbf{61.91\%} & \textbf{81.54\%} & \textbf{86.12\%} & \textbf{0.7042} & \textbf{0.5298} \\ \cline{2-7} 
                         & \Name{}+    & \textbf{64.16\%} & \textbf{83.79\%} & \textbf{88.37\%} & \textbf{0.7265} & \textbf{0.5524} \\ \hline
\multirow{9}{*}{ST}      & BugLocator & 39.66\% & 68.34\% & 77.93\% & 0.5183 & 0.3954 \\ \cline{2-7} 
                         & BLUiR      & 29.21\% & 55.86\% & 65.99\% & 0.4065 & 0.3073 \\ \cline{2-7} 
                         & BRTracer   & 46.16\% & 76.12\% & 84.22\% & 0.5897 & 0.4392 \\ \cline{2-7} 
                         & AmaLgam    & 29.32\% & 55.97\% & 66.31\% & 0.4076 & 0.3084 \\ \cline{2-7} 
                         & BLIA       & 46.06\% & 74.20\% & 83.26\% & 0.5797 & 0.4574 \\ \cline{2-7} 
                         & BLIZZARD   & 36.46\% & 60.98\% & 69.72\% & 0.4663 & 0.3486 \\ \cline{2-7} 
                         & LR         & 32.62\% & 62.37\% & 75.16\% & 0.4513 & 0.3440 \\ \cline{2-7} 
                         & \Name{}    & \textbf{51.71\%} & \textbf{79.64\%} & \textbf{85.61\%} & \textbf{0.6391} & \textbf{0.4802} \\ \cline{2-7} 
                         & \Name{}+    & \textbf{52.45\%} & \textbf{80.38\%} & \textbf{86.35\%} & \textbf{0.6461} & \textbf{0.4872} \\ \hline
\multirow{9}{*}{NL}      & BugLocator & 24.66\% & 51.16\% & 64.80\% & 0.3632 & 0.2560 \\ \cline{2-7} 
                         & BLUiR      & 24.95\% & 52.42\% & 62.96\% & 0.3668 & 0.2615 \\ \cline{2-7} 
                         & BRTracer   & 28.05\% & 54.84\% & 67.21\% & 0.3967 & 0.2736 \\ \cline{2-7} 
                         & AmaLgam    & 25.05\% & 52.51\% & 63.15\% & 0.3679 & 0.2618 \\ \cline{2-7} 
                         & BLIA       & 17.50\% & 45.94\% & 58.12\% & 0.2875 & 0.2032 \\ \cline{2-7} 
                         & BLIZZARD   & 21.47\% & 43.91\% & 55.22\% & 0.3133 & 0.2117 \\ \cline{2-7} 
                         & LR         & 23.99\% & 60.13\% & 76.65\% & 0.3507 & 0.2575 \\ \cline{2-7} 
                         & \Name{}     & \textbf{32.24\%} & \textbf{61.32\%} & \textbf{72.63\%} & \textbf{0.4608} & \textbf{0.3228} \\ \cline{2-7} 
                         & \Name{}+    & \textbf{37.43\%} & \textbf{64.51\%} & \textbf{75.82\%} & \textbf{0.4923} & \textbf{0.3540} \\ \hline
\multirow{9}{*}{ALL}     & BugLocator & 44.33\% & 70.18\% & 78.53\% & 0.5534 & 0.4161 \\ \cline{2-7} 
                         & BLUiR      & 36.93\% & 64.98\% & 73.74\% & 0.4868 & 0.3769 \\ \cline{2-7} 
                         & BRTracer   & 48.75\% & 73.41\% & 81.30\% & 0.5947 & 0.4408 \\ \cline{2-7} 
                         & AmaLgam    & 37.18\% & 65.21\% & 74.06\% & 0.4892 & 0.3791 \\ \cline{2-7} 
                         & BLIA       & 41.51\% & 69.54\% & 78.77\% & 0.5326 & 0.4151 \\ \cline{2-7} 
                         & BLIZZARD   & 36.11\% & 61.52\% & 70.48\% & 0.4690 & 0.3465 \\ \cline{2-7} 
                         & LR         & 40.96\% & 67.62\% & 77.97\% & 0.5212 & 0.3865 \\ \cline{2-7} 
                         & \Name{}     & \textbf{55.21\%} & \textbf{77.56\%} & \textbf{83.59\%} & \textbf{0.6494} & \textbf{0.4841} \\ \cline{2-7} 
                         & \Name{}+    & \textbf{57.98\%} & \textbf{80.33\%} & \textbf{86.36\%} & \textbf{0.6770} & \textbf{0.5118} \\ \hline
\end{tabular}
\vspace*{-15pt}
\end{table}

For PE (Programming Entities), considered ideal for IRFL, 
all techniques demonstrate better performance compared to ST and NL. 
In particular, BRTracer achieves a high MRR of 0.6508 and a high MAP of 0.4875. 
However, \Name{} surpasses BRTracer with a Top1 accuracy of 61.91\%, Top5 accuracy of 81.54\%, Top10 accuracy of 86.12\%, 
an MRR of 0.7042, and a MAP of 0.5298, outperforming BRTracer's MRR and MAP by 8\% and 9\%, respectively. 
These results indicate that \Name{} effectively reduces noise and enhances the performance of IRFL.

For bug reports that include stack traces (ST), 
BRTracer introduced a scoring method based on the rank position of the entries in the stack traces. 
BLIA has since adopted this method from BRTracer. 
As the results indicate, this approach effectively computes the scores, leading their IRFL performance to surpass other state-of-the-art techniques. 
\Name{} leverages GPT-4’s strength which lies in its contextual comprehension to analyze stack traces. 
By analyzing the entire stack trace, the model can discern patterns or anomalies. 
For instance, it can identify if a specific method call consistently precedes an error, suggesting potential trouble spots. 
For ST, \Name{} achieves an MRR and MAP of 0.6391 and 0.4802, outpacing BRTracer by 8\% and 9\%, 
respectively, and outpacing BLIA  by 10\% and 5\%, respectively.

For bug reports described only in plain text without programming entities or stack traces (NL) (Table~\ref{compare_stoa}), 
BLIZZARD proposed an expansion method via pseudo-relevance feedback; other state-of-the-art techniques lack specific methods. 
Results suggest that while BLIZZARD's expansion may be beneficial in certain scenarios, 
on average, it is less efficient than methods treating bug reports as queries alone. 
This inefficiency may arise from improper expansions introducing extraneous information. 
\Name{} leverages GPT-4's expansive knowledge and contextual understanding to expand queries. 
\Name{} achieves performance metrics of 32.24\% for Top1, 61.32\% for Top5, 72.63\% for Top10, an MRR of 0.4608, 
and a MAP of 0.3228  on ${NL}$, surpassing all competitors. 
This suggests that \Name{} adeptly infers and recommends classes or components associated with the given bug report.

After the performance evaluation of \Name{}, it was observed that 986 out of 6,340 bug reports did not rank in the generated list. 
This discrepancy can be attributed to the inadequate or misleading information contained within these reports, such as incorrect references to non-buggy classes. 
To address this issue, \Name{}+ was employed to perform query reformulation on these problematic bug reports, aiming to refine the queries and improve the accuracy of the retrieval results.
After the reformulation process executed by \Name{}+, we observed that 156 out of the 986 previously unranked bug reports were successfully included in the ranking list. This improvement indicates that the query reformulation capability of \Name{}+ can effectively rectify issues related to inadequate or misleading information in bug reports, thereby enhancing the overall accuracy and reliability of the fault localization process.

\vspace*{3pt}
\begin{tcolorbox}
\stepcounter{findingsCounter}
{\bf RQ1 summary:} {
The evaluation results demonstrate that \Name{} and its advanced version, \Name{}+, surpass six leading IRFL techniques. \Name{}+ consistently delivers exceptional results, outdoing BRTracer by 9\% in MRR and 10\% in MAP. \Name{} also excels in analyzing stack traces ($ST$), leveraging GPT-4’s contextual understanding to outperform competing techniques. In handling plain text descriptions ($NL$), \Name{} shows remarkable proficiency, effectively inferring and recommending relevant components. With \Name{}+, which incorporates dynamic user-driven query reformulation, the system further enhances its performance, overcoming real-world challenges and surpassing existing state-of-the-art solutions.
}
\end{tcolorbox}

\subsubsection{RQ2: How does the effectiveness of our LtR model compare to state-of-the-art IR models?} 


To answer this research question, we first evaluate how each feature impacts our Learning-to-Rank model and determine the best combination of features. Then, we compare our optimized Learning-to-Rank model with state-of-the-art retrieval models.

\noindent
\textbf{Features Selection.} With a set of 7 features (Section~\ref{fs}), 
we conducted a comprehensive evaluation of all possible combinations to ascertain the most effective feature set. Table~\ref{features} presents the outcomes, highlighting the optimal feature combinations tailored to each category of bug reports. 
Given that our queries may not always align with class names and method names, 
these features could sometimes result in zero scores. Similarly, call graph and class hierarchy features and other features may not always be applicable, 
but textual similarity between bug reports and source files generally provides a consistent score. Thus, textual similarity serves as a foundational feature, complemented by the other features.


\begin{table}[h]\footnotesize
\setlength{\tabcolsep}{6pt}
\caption{Comparing with Different Features.} \label{features}
\begin{tabular}{|l|cc|cc|cc|}
\hline
\multirow{2}{*}{\textbf{Features}} & \multicolumn{2}{c|}{\textbf{PE}}                       & \multicolumn{2}{c|}{\textbf{ST}}                       & \multicolumn{2}{c|}{\textbf{NL}}                       \\ \cline{2-7} 
                                   & \multicolumn{1}{c|}{MRR}             & MAP             & \multicolumn{1}{c|}{MRR}             & MAP             & \multicolumn{1}{c|}{MRR}             & MAP             \\ \hline \hline
TS                                 & \multicolumn{1}{c|}{0.5785}          & 0.4356          & \multicolumn{1}{c|}{0.5316}          & 0.3997          & \multicolumn{1}{c|}{0.3273}          & 0.2340          \\ \hline
TS+CL                              & \multicolumn{1}{c|}{0.6654}          & 0.4992          & \multicolumn{1}{c|}{0.6055}          & 0.4454          & \multicolumn{1}{c|}{\textbf{0.4124}} & \textbf{0.3127} \\ \hline
TS+CL+CG                           & \multicolumn{1}{c|}{\textbf{0.6749}} & \textbf{0.5056} & \multicolumn{1}{c|}{\textbf{0.6333}} & \textbf{0.4644} & \multicolumn{1}{c|}{0.3528}          & 0.2604          \\ \hline
ALL                                & \multicolumn{1}{c|}{0.6686}          & 0.5014          & \multicolumn{1}{c|}{0.6152}          & 0.4537          & \multicolumn{1}{c|}{0.3528}          & 0.2604          \\ \hline
\end{tabular}
\begin{flushleft}
\scriptsize
\textbf{TS}: Text Similarity, \textbf{CL}: Class Match, \textbf{CG}: Call Graph, \textbf{ALL}: All features combined.
\end{flushleft}
\end{table}

The results show that for bug reports with programming entities (PE) and stack traces (ST), the queries yield accurate results, with class match and call graph scores proving highly effective. The combination of text similarity (TS), class match (CL), and call graph (CG) delivers the best performance, while adding more features can increase false positives and reduce effectiveness. For natural language (NL) reports, \Name{}'s GPT-4-driven queries suggest multiple root causes, mixing true and false positives. In this case, combining class matching with textual similarity works best, as incorporating call graph scores tends to lower overall accuracy due to increased false positives.

\noindent
\textbf{Comparison with State-of-the-Art Retrieval Models.}  
We evaluated our learning-to-rank model against leading retrieval models, including BugLocator, BLUiR, BRTracer, AmaLgam, BLIA, and BLIZZARD, using a randomized subset of the dataset and focusing on MRR and MAP. The results, shown in Figure~\ref{model}, highlight our model's superior performance and its alignment with the specialized nature of our queries, demonstrating its suitability for retrieval tasks.

\vspace*{3pt}
\begin{tcolorbox}
\stepcounter{findingsCounter}
{\bf RQ2 summary:} {
We identified the best feature combinations for different bug report categories. For PE and ST, the optimal mix included Text Similarity, Class Match, and Call Graph, while for NL reports, Text Similarity and Class Match were sufficient. Our Learning-to-Rank model outperformed traditional IR models, demonstrating its effectiveness and superior performance.
}
\end{tcolorbox}

\subsubsection{RQ3: How effective are our queries in IRFL?} 
To answer to this question, we compared our queries against others commonly used in IRFL, then examined the effects of conversational query reformulation and the influence of one-shot/zero-shot learning on our query performance.

\noindent
\textbf{Compared with Different Queries.} 
To determine the effectiveness of our queries, we benchmarked them against common IRFL practices, which use entire bug reports (BR) as queries, and the BLIZZARD approach, which generates queries using a graph-based method. 
To ensure consistency in our evaluation, we perform all queries using our learning-to-rank model.
As shown in Table~\ref{queries}, our queries surpass both BR and BLIZZARD for PE and ST categories, signifying that our method enables GPT-4 to more accurately pinpoint the true root causes within bug reports. For NL categories, despite BLIZZARD's use of Pseudo-relevance feedback, our queries still perform better, demonstrating the efficacy of leveraging GPT-4's knowledge to refine the queries.

\begin{table}[h]\footnotesize
\caption{Comparing with Different Queries.} \label{queries}
\begin{tabular}{|l|cc|cc|cc|}
\hline
\multirow{2}{*}{\textbf{Features}} & \multicolumn{2}{c|}{\textbf{PE}} & \multicolumn{2}{c|}{\textbf{ST}} & \multicolumn{2}{c|}{\textbf{NL}} \\ \cline{2-7} 
                                   & \multicolumn{1}{c|}{MRR}  & MAP  & \multicolumn{1}{c|}{MRR}  & MAP  & \multicolumn{1}{c|}{MRR}  & MAP  \\ \hline \hline
BR                                 & \multicolumn{1}{c|}{0.5172}     & 0.3804     & \multicolumn{1}{c|}{0.3931}     & 0.2702     & \multicolumn{1}{c|}{0.3599}     &  0.2051    \\ \hline
BLIZZARD                           & \multicolumn{1}{c|}{0.5161}     & 0.3814     & \multicolumn{1}{c|}{0.4018}     & 0.2765     & \multicolumn{1}{c|}{0.3660}     & 0.2574     \\ \hline
\Name{}           & \multicolumn{1}{c|}{\textbf{0.7680}}     &  \textbf{0.5816}    & \multicolumn{1}{c|}{\textbf{0.6554}}     &  \textbf{0.4841}    & \multicolumn{1}{c|}{\textbf{0.4867}}     &  \textbf{0.3030}    \\ \hline
\end{tabular}
\end{table}

\begin{figure}[htbp]
	\centering
	\begin{minipage}{0.45\linewidth}
		\centering
		\includegraphics[width=1\linewidth]{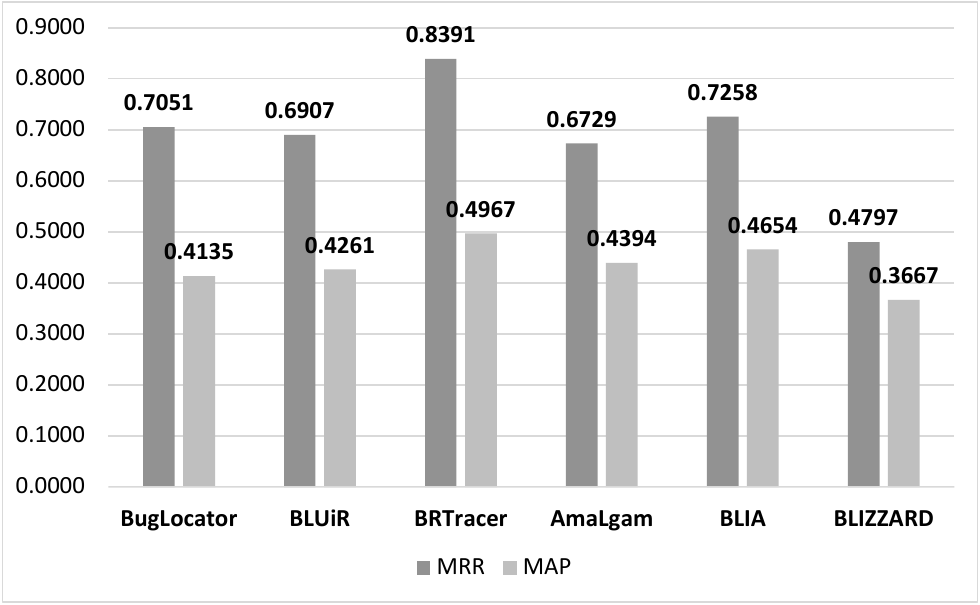}
		\subcaption{Retrieval models.}
		\label{model}
	\end{minipage}
	\begin{minipage}{0.45\linewidth}
		\centering
		\includegraphics[width=1\linewidth]{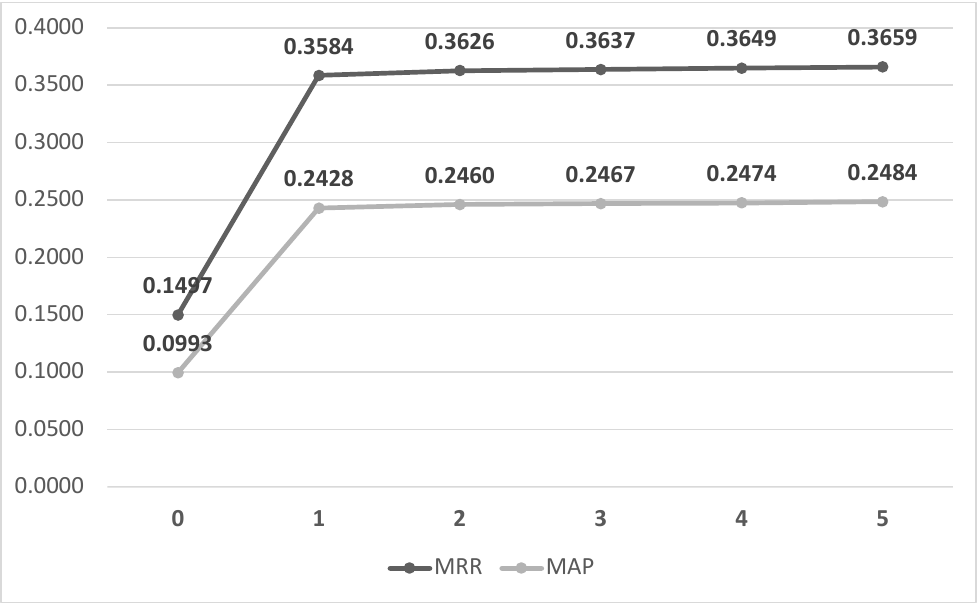}
		\subcaption{Conversation cycles.}
		\label{cycles}
	\end{minipage}
	\caption{The impact of various components on \Name{}.}\label{components}
\end{figure}

\noindent
\textbf{Conversation Cycles.} 
To assess the efficiency of our query reformulation process, we analyzed 1,743 bug reports from Dataset requiring reformulation. Setting a limit of 5 conversation cycles, we recorded MRR and MAP performance. Figure~\ref{cycles} reveals that peak performance is typically reached in the first cycle, averaging an MRR of 0.3584 and a MAP of 0.2428. Our conversation-based reformulation efficiently responds to initial user feedback, resulting in significant query improvement. Subsequent cycles provide marginal enhancements, suggesting rapid peak performance attainment. Hence, we implemented \Name{}+'s one conversation cycle.

\noindent
\textbf{0/1 Shot.} 
0/1-shot learning assesses the model's adaptability across diverse tasks without extensive fine-tuning. Our experiment, conducted on a randomized dataset subset, yielded results in Table~\ref{shots}. The 1-shot approach outperformed 0-shot in PE, ST, and NL bug report categories. While 1-shot showed no significant difference from 0-shot in extracting programming entities, tasks in ST and NL resembled 0-shot for GPT. Notably, the 1-shot method excelled in ST and NL, leveraging initial context for accurate queries. The 0-shot occasionally produced non-optimally formatted output, requiring human effort to reformat queries. Consequently, we adopted the 1-shot approach for \Name{} and \Name{}+.

\begin{table}[]\footnotesize
\caption{Comparison between 0-shot and 1-shot.} \label{shots}
\begin{tabular}{|c|l|c|c|c|c|c|}
\hline
                     & shots  & Top1    & Top5    & Top10   & MRR    & MAP    \\ \hline
\multirow{2}{*}{PE}  & 0-shot & 60.32\% & \textbf{82.02\%*} & \textbf{84.79\%*} & 0.6958 & 0.4999 \\ \cline{2-7} 
                     & 1-shot & \textbf{61.17\%*} & 81.38\% & 83.51\% & \textbf{0.7014*} & \textbf{0.5024*} \\ \hline
\multirow{2}{*}{ST}  & 0-shot & 34.34\% & 78.85\% & 87.91\% & 0.5241 & 0.4044 \\ \cline{2-7} 
                     & 1-shot & \textbf{59.62\%} & \textbf{84.89\%} & \textbf{87.91\%} & \textbf{0.7092} & \textbf{0.5545} \\ \hline
\multirow{2}{*}{NL}  & 0-shot & 14.89\% & 37.81\% & 48.48\% & 0.2453 & 0.1567 \\ \cline{2-7} 
                     & 1-shot & \textbf{20.16\%} & \textbf{47.96\%} & \textbf{57.71\%} & \textbf{0.3158} & \textbf{0.2020} \\ \hline
\multirow{2}{*}{ALL} & 0-shot & 28.43\% & 47.49\% & 52.44\% & 0.3641 & 0.2599 \\ \cline{2-7} 
                     & 1-shot & \textbf{35.17\%} & \textbf{56.07\%} & \textbf{61.33\%} & \textbf{0.4423} & \textbf{0.3148} \\ \hline
\end{tabular}
\end{table}

\vspace*{3pt}
\begin{tcolorbox}
\stepcounter{findingsCounter}
{\bf RQ3 summary:} {
Our queries outperformed standard IRFL queries in PE and ST categories and remained effective in NL categories. Conversational query reformulation led to quick performance gains, with one-shot learning proving especially beneficial for ST and NL bug reports.
}
\end{tcolorbox}
\section{Related Work}

\noindent\textbf{Large Language Model.}
In recent years, Large Language Models (LLMs)~\cite{vaswani2017attention, devlin2018bert, radford2019language} have significantly advanced the field of natural language processing by offering transformative capabilities in 
understanding and generating human-like text. LLMs, such as OpenAI's GPT~\cite{GPT-4} (Generative Pre-trained Transformer) series, are trained on extensive corpora of textual 
data and utilize deep learning techniques, particularly transformer architectures, to capture intricate patterns and nuances in language. 
Their applications span a wide range of tasks, from text completion and translation to more complex endeavors like question-answering, 
summarization, and even code generation.

Recently, LLMs have been widely applied to software engineering tasks, such as code generation, code completion~\cite{chen2021evaluating, doderlein2022piloting, izadi2022codefill}, code search~\cite{salza2022effectiveness, shi2022cross}, 
and program repair~\cite{xia2023conversational, yuan2022circle}. Ciborowska et al.~\cite{ciborowska2022fast} presented an approach for automatically retrieving bug-inducing changesets for newly reported bugs, 
using the popular BERT model to more accurately match the semantics in the bug report text to the inducing changeset. 
They also proposed using data augmentation (DA)~\cite{ciborowska2023too} to create new, realistically looking bug reports that can significantly increase the size of the training set. 

More recently, Kang et al.~\cite{kang2024quantitative} proposed AutoFL, a fault localization method based on LLMs that generates both fault locations and natural language explanations for bug reports. By effectively leveraging LLMs to provide rationales alongside fault predictions, AutoFL improves developer trust in the recommendations. Additionally, Yoo et al.~\cite{qin2024agentfl} introduced AgentFL, a multi-agent fault localization system built on ChatGPT. By simulating a three-step debugging process (comprehension, navigation, and confirmation), AgentFL employs diverse agents with specialized expertise and utilizes strategies such as test behavior tracking and multi-round dialogue to localize faults effectively. 

However, these works do not focus on IR-based fault localization. Ciborowska's work requires training a model, while AutoFL and AgentFL rely on test coverage information. In contrast, our approach only relies on bug reports and simply prompts the model.

\noindent\textbf{Information Retrieval-based Fault Localization.} 
IR-based fault localization is an effective technique for identifying bugs in software systems. 
Automated methods are particularly valuable in this context to circumvent the time-consuming and error-prone nature of manual code inspection. 
Gay et al.~\cite{gay2009use} proposed a relevance feedback mechanism to enhance the vector space model (VSM) used in retrieval processes. 
Zhou et al.~\cite{buglocator} further improved the VSM model by incorporating document length, leading to the development of the revised VSM (rVSM). 
Saha et al.~\cite{bluir} utilized the Eclipse JDT to parse the source code's abstract syntax tree (AST) and extract information from four types of code entities, 
thereby augmenting the efficiency of IR-based fault localization. 
Additionally, Wang et al.~\cite{amalgam} introduced an approach that leverages version history information to refine IR-based fault localization techniques. 
In our research, we have conducted a comparative analysis of our approach with these established localization techniques. 
The results demonstrate that our approach can outperform all these methods, showcasing its effectiveness in the realm of fault localization. 

\section{Threats to Validity}

The primary threat to the external validity of this study involves its generalizability across domains, programming languages, project sizes, and user expertise levels. To address these concerns, we employed a dataset comprising 46 diverse projects, enhancing the approach's applicability. We also mitigated user expertise variability by implementing a structured feedback template, minimizing the reliance on users' technical articulation and familiarity with the software system.



The primary threat to internal validity stems from reliance on the GPT model for bug report analysis, which may be limited by its understanding of complex technical contexts. The accuracy of generated queries and the effectiveness of query reformulation hinge on GPT's interpretation of user feedback, introducing the risk of inaccurate adjustments due to misinterpretation or ambiguity. To address these concerns, we emphasize GPT's advanced capabilities and proven efficacy in diverse applications. Recognized as one of the most powerful language models, GPT has demonstrated robust performance across numerous domains, accurately interpreting and processing complex language constructs in various tasks.



\section{Conclusion}


Our approach, named \Name{}, categorizes bug reports into programming entities, stack traces, and pure text, tailoring query construction strategies. Introducing \Name{}+, a novel user feedback-driven approach.
Moreover, to optimize query effectiveness further, we have incorporated a learning-to-rank model that utilizes crucial features including class name match scores and call graph scores.
We demonstrated significant outperformance of existing IRFL techniques in our evaluation of 46 projects and 6,340 bug reports, achieving an MRR of 0.6770 and a MAP of 0.5118. 
This marks a notable advancement in software engineering. Future work will refine GPT-based fault localization, with a focus on enhancing method-level capabilities.


\newpage
\bibliographystyle{ACM-Reference-Format}
\bibliography{sample-base}

\end{document}